\documentclass[pra,twocolumn,showpacs,preprintnumbers,amsmath,amssymb]{revtex4-1}

\usepackage{graphicx}
\usepackage{dcolumn}
\usepackage{bm}
\usepackage{epsfig}
\usepackage{amsmath}
\usepackage{amssymb}
\usepackage{color}
\usepackage{bbm}
\usepackage[caption=false]{subfig}
\usepackage{placeins}

\usepackage[colorlinks=true,citecolor=blue,linkcolor=blue,urlcolor=blue]{hyperref}
\usepackage{cancel}
\newcommand{\bra}[1]{\langle#1|}
\newcommand{\ket}[1]{|#1\rangle}

\newcommand{\braOket}[3]{\langle#1|#2|#3\rangle}
\newcommand{\erw}[1]{\langle#1\rangle}

\newcommand{\abs}[1]{\lvert#1\rvert}

\def\equationautorefname~#1\null{%
  Eq.~#1\null
}
\def\figureautorefname~#1\null{%
	Fig.~#1\null
}
\graphicspath{{./figs/}}

\begin{document}

\title{Active energy transport and the role of symmetry breaking in microscopic power grids}

\author{Julian Huber}
\affiliation{Vienna Center for Quantum Science and Technology,
Atominstitut, TU Wien, 1040 Vienna, Austria}
\author{Peter Rabl}
\affiliation{Vienna Center for Quantum Science and Technology,
Atominstitut, TU Wien, 1040 Vienna, Austria}

\date{\today}

\begin{abstract}
We study the  transfer of energy through a network of coupled oscillators, which represents a minimal microscopic power grid connecting multiple active quantum machines. We evaluate the resulting energy currents in the macroscopic, the thermal and the quantum regime and describe how transport is affected by the competition between coherent and incoherent processes and nonlinear saturation effects.  Specifically, we show that the transfer of energy through such networks is strongly influenced by a non-equilibrium phase transition between a noise-dominated and a coherent transport regime. This transition is associated with the formation and breaking of spatial symmetries and is identified as a generic feature of active networks.  Therefore, these findings have  important practical consequences for the distribution of energy over coherent microwave, optical or phononic channels, in particular close to or at the quantum limit. 
\end{abstract}

\maketitle
\section{Introduction}
Motivated by fundamental thermodynamical considerations as well as potential practical implications, there has recently been a growing interest in the performance of microscopic generators, engines or refrigerators, which may even be realized with single quantum systems \cite{Gelbwaser2015,Vinjanampathy2016,Goold2016,Alicki2018,Millen2016}. However, while many theoretical  \cite{Scovil1959,Geva1994,Scully2004,Kieu2004,Quan2007,Linden2010,Abah2012,Gelbwaser2013,Gallego2014,Zhang2014,Bergenfeldt2014,Brunelli2015,Elourad2015,Dechant2015,Mari2015,Li2017} and first experimental \cite{Hugel2002,Steeneken2011,Blickle2012,Brantut2013,Thierschmann2015,Rossnagel2016,Schmidt2018,Klaers2017} studies of individual quantum machines have already been performed, there is still little known about interfacing multiple such devices. For example, how can the energy output of a microscopic generator be efficiently delivered to a microscopic engine and how will even larger networks of active quantum machines behave? Compared to conventional transmission lines for electric power, energy can be distributed at the microscale via highly coherent nanophotonic, microwave or micromechanical channels, while at the same time thermal and quantum fluctuations become important. Thus,  the flow of energy through such microscopic networks can differ strongly from the usual Ohm's or Fourier's law.

\begin{figure}
		\includegraphics[width=\columnwidth]{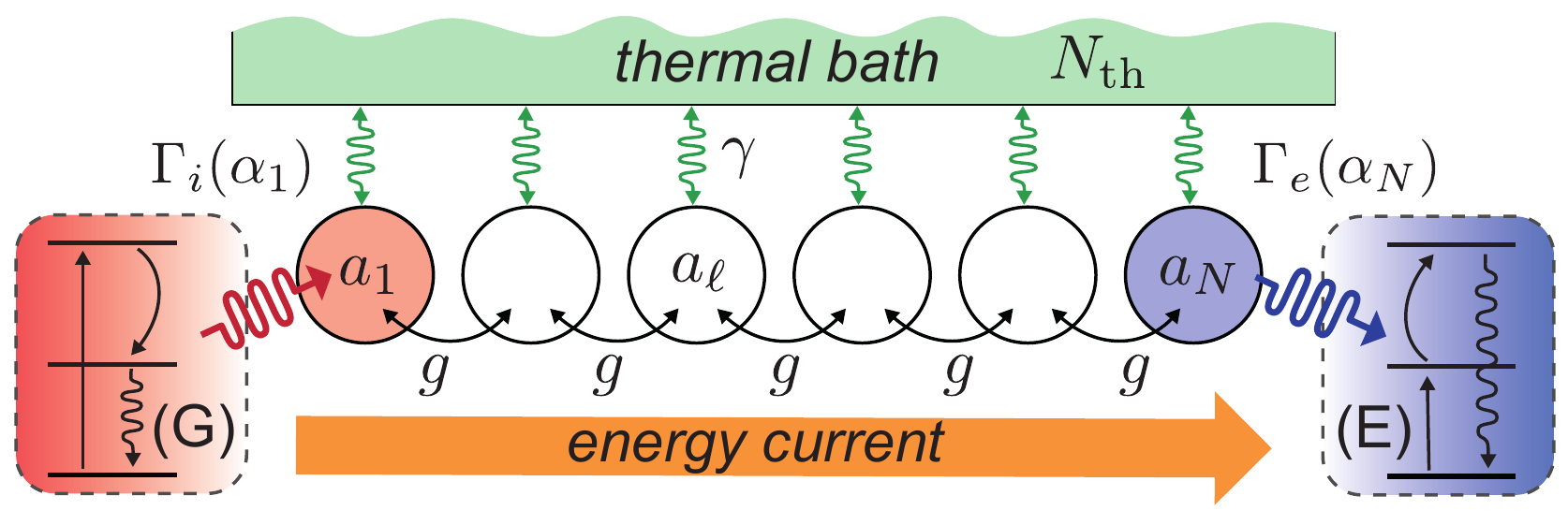}
	\caption{Sketch of a microscopic power grid realized by an array of coupled harmonic oscillators. Energy is injected at one end by a microscopic generator (G) with rate $\Gamma_i$ and extracted at the other end by a microscopic engine (E) with rate  $\Gamma_e$. Both processes are energy-dependent and saturate above a characteristic amplitude  $\sqrt{n_0}$. All oscillators are coupled weakly to a thermal environment. See text for more details.}
	\label{fig:energytransportthermal}
\end{figure}

In this work we analyze the energy transfer between two quantum machines, which are connected through  a `power grid' of coupled harmonic oscillators, as depicted in Fig.~\ref{fig:energytransportthermal}.  Such oscillator chains have been considered in the past to study the passive transfer of heat between thermal reservoirs \cite{Rieder1967,Lin2011,Asadian2013,Bermudez2013,Motz2017}, often with the goal to investigate the transition from ballistic transport to Fourier's law by adding additional dephasing mechanisms. Here we are interested in a different scenario where energy is injected at one end by a microscopic generator and extracted at the other end of the chain by a microscopic engine. 
Such quantum machines are active devices, meaning that they are (i) operated under non-equilibrium conditions and (ii) characterized by a maximal rate at which energy quanta can be emitted or absorbed. These properties make the problem of active energy transfer very distinct from the study of heat or electric transport between large passive reservoirs. By varying a single saturation parameter,  we can tune the degree of microscopicity of our network and investigate the resulting transport phenomena in the macroscopic, in the thermal and deep in the quantum regime.

Despite its conceptual simplicity, this setting already exhibits several surprising features, which will be relevant as well for the operation of more general  networks. Most importantly,  we identify two basic modes of transport, which are separated by a sharp phase transition and differ strongly in their noise characteristics. Such sharp transitions are unfamiliar in heat transport---neither in the ballistic nor in the Fourier regime---where mean currents and fluctuations vary smoothly as a function of the energy injection and extraction rates. The observed transition is accompanied by a breaking of the parity symmetry of the steady-state energy distribution along the channel. This  effect is closely related to the phenomenon of $\mathcal{PT}$-symmetry breaking \cite{Bender1998,Hassan2015,Kepesidis2016,Ge2016,Assawaworrarit2017,Ganainy2018} in systems with exactly balanced gain and loss. Importantly, here we show that the formation and breaking of spatial symmetries plays a much more fundamental  role for energy transport in coherent networks,  even in systems where such symmetries are not reflected in the underlying equations of motion. Therefore, these findings have direct practical consequences for various energy-distribution schemes at the microscopic level, but also reveal an interesting fundamental connection between symmetry-breaking, non-equilibrium phase transitions and the operation of networks of coupled quantum machines.

\section{Model}
We consider a chain of $N\geq 2$ coupled harmonic oscillators, as schematically shown in Fig. \ref{fig:energytransportthermal}. The oscillators have a frequency $\omega_0$ and they are coupled to their neighbors with strength $g$. Energy is injected at the first site with a rate $\Gamma_i$ and extracted at the other end with rate $\Gamma_e$. In addition, all oscillators are weakly coupled to a thermal environment at temperature $T$. In the frame rotating with  $\omega_0\gg g,\Gamma_{i,e}$, the whole network is described by a master equation for the density operator $\rho$, 
\begin{equation}
\begin{split}
\dot \rho=&  - \frac{i}{\hbar} [H_{g},\rho] +\Gamma_{i} \mathcal{D}[A_1^\dag]\rho+\Gamma_{e}\mathcal{D}[A_{N}]\rho \\
&+ \sum_{\ell=1}^{N} \gamma (N_{\rm th}+1) \mathcal{D}[a_{\ell}]\rho+\gamma N_{\rm th} \mathcal{D}[a_{\ell}^{\dagger}]\rho,
\end{split}
\label{eq:mastereq}
\end{equation}
where $a_\ell$ ($a_\ell^\dag$) are the annihilation (creation) operators for each oscillator and $\mathcal{D}[a] \rho \equiv  a \rho a^{\dagger} - (a^{\dagger} a \rho- \rho a^\dagger a)/2$.
In Eq.~\eqref{eq:mastereq}, $H_g= -\frac{\hbar g}{2} \sum_{\ell=1}^{N-1}(a_{\ell}^\dagger a_{\ell+1} +{\rm H.c.})$ describes the coherent exchange of energy along the chain, while the second and the third term model the incoherent pump and dissipation processes, respectively. 

As mentioned in the introduction, in this work we are interested in active energy transport, where source and drain are represented by driven few-level quantum systems. However, to keep our analysis on a general level and to avoid details of specific implementations, we simply mimic the main characteristics of such microscopic generators and engines (providing gain, being saturable) by introducing in Eq.~\eqref{eq:mastereq} the nonlinear jump operators $A_{\ell=1,N}=f(a_{\ell}^\dagger a_{\ell}) a_{\ell}$. Here the cutoff function $f(x)$, where $f(0)=1$ and $f(x\gg n_0)\rightarrow0$, accounts for the fact that both the injection as well as the extraction of energy saturate above a characteristic occupation number $n_0$. By changing this saturation parameter, we can tune the degree of microscopicity of the network without changing any other properties of the system. For concreteness, we will focus here on the cutoff function
\begin{equation}\label{eq:f}
f(a^\dagger a)=\frac{1}{(1+a^\dag a/n_0)},
\end{equation} 
which reproduces the saturation dependence of driven three-level generators and engines~\cite{Kepesidis2013}, as depicted in Fig. \ref{fig:energytransportthermal}. However, none of the central conclusions of this work depends on this assumption and different shapes of $f(x)$ can be used to model other realizations of quantum machines discussed in the literature~\cite{Scovil1959,Geva1994,Scully2004,Kieu2004,Quan2007,Linden2010,Abah2012,Gelbwaser2013,Gallego2014,Zhang2014,Bergenfeldt2014,Brunelli2015,Elourad2015,Dechant2015,Mari2015,Li2017}.

Finally, the second line of Eq.~\eqref{eq:mastereq} describes the coupling of each oscillator to a local thermal bath, where $N_{\rm th}=(e^{\hbar\omega_0/k_BT}-1)^{-1}$ is the equilibrium occupation number and $\gamma$ the damping rate, which we assume to be much smaller than $\Gamma_i$ and $\Gamma_e$. Note that the use of local jump operators in Eq.~\eqref{eq:mastereq} is justified by the assumption that $\omega_0$ is large compared to both the coherent intra-system coupling $g$ and the dissipation rates \cite{Plenio2018,hofer2017}. Typical systems which can be used to implement this model include coupled nanomechanical resonators \cite{Hatanaka2014,Huang2016,Cha2018,Patel2018},  linear chains of trapped ions \cite{Lin2011,Bermudez2013,Haeffner2014} or arrays of coupled $LC$ oscillators \cite{Underwood2012,Houck2017,Mirhosseini2018}. For all those platforms various techniques for engineering local gain and loss processes at the quantum level are already experimentally available~\cite{Aspelmeyer2014, Leibfried2003,Gu2017}.

For most parts of the following discussion we will be interested in the regime $n_0\gg1$,  where Eq.~\eqref{eq:mastereq} can be mapped onto a Fokker-Planck equation for the Glauber-Sudarshan P-distribution $P(\{\alpha_\ell\},t)$ \cite{WallsMilburn,GardinerZoller} (for details see  Appendix~\ref{app:FP}). This distribution can be sampled numerically by integrating the corresponding stochastic Ito equations for the amplitudes $\alpha_\ell$ \cite{Gardiner},  
\begin{eqnarray}
\dot{\alpha_{1}}&=&\frac{\Gamma_{i}(\alpha_{1})-\gamma}{2}\alpha_{1} +i \frac{g}{2} \alpha_{2}+\sqrt{D_{\rm th}\!+\!\Gamma_{i}(\alpha_{1})} \xi_{1}(t),\label{eq:itoeq1}\\
\dot{\alpha_{\ell}}&=&- \frac{\gamma}{2} \alpha_{\ell}+i \frac{g}{2}\left( \alpha_{\ell-1} + \alpha_{\ell+1}\right) + \sqrt{D_{\rm th}} \xi_{\ell}(t),\label{eq:itoeq2}\\
\dot{\alpha_{N}}&=&-\frac{\Gamma_{e}(\alpha_{N})+\gamma}{2} \alpha_{N} + i \frac{g}{2} \alpha_{N-1}+ \sqrt{D_{\rm th}} \xi_{N}(t).
\label{eq:itoeq3}
\end{eqnarray}
Here  $\Gamma_{i,e}(\alpha)= \Gamma_{i,e} f^2(|\alpha|^2)$ and $D_{\rm th}=\gamma N_{\rm th}$ is the thermal diffusion rate. The $\xi_{\ell}(t)$ are white noise processes, which satisfy $\erw{\xi_{\ell}^*(t)\xi_{\ell'}(t')}=\delta_{\ell\ell'}\delta (t-t') $. We are primarily interested in the steady-state energy current $\langle J_\ell\rangle = i\frac{g}{2} \erw{a_{\ell}^{\dagger} a_{\ell-1}-a_{\ell-1}^\dagger  a_{\ell}}= g {\rm Im} \langle\langle \alpha_{\ell-1}^* \alpha_{\ell}\rangle\rangle$, which can be obtained from the longtime average over many trajectories, denoted by $\langle\langle \cdot \rangle\rangle$. In the regime of interest, $\gamma\rightarrow 0$, the average current is approximately constant throughout the chain and we can drop the index $\ell$.

\begin{figure}
	\includegraphics[width=1\columnwidth]{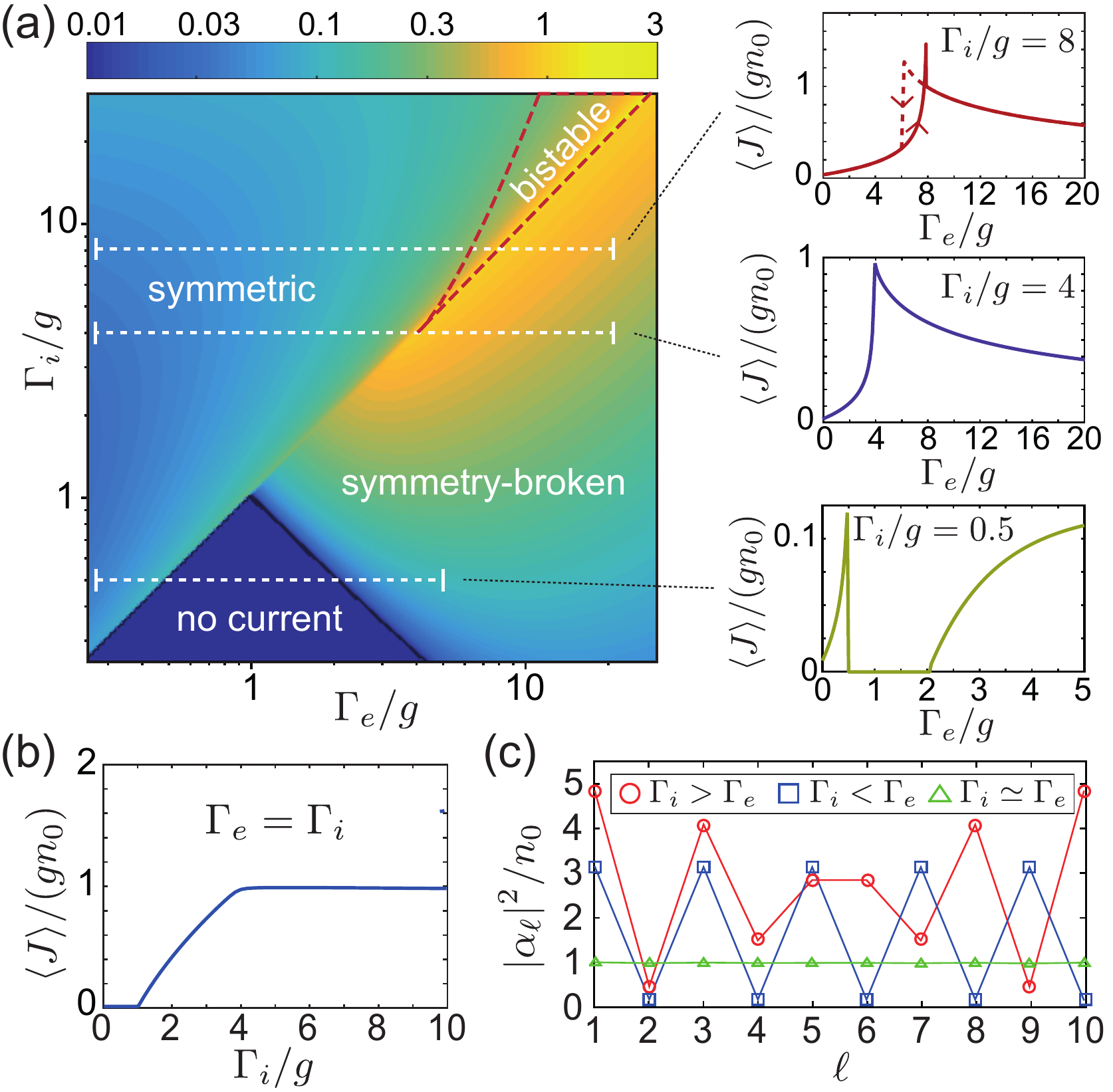}
	\caption{(a) Dependence of the normalized steady-state energy current $\erw{J}/(gn_0)$  on the energy injection and extraction rates, $\Gamma_i$ and $\Gamma_e$, for a chain of $N=10$ oscillators. (b) Plot of the average current $\erw{J}$ under fully symmetric conditions, $\Gamma_e=\Gamma_i$. (c) The steady-state occupation numbers $|\alpha_\ell^0|^2$ of the whole chain are plotted in the symmetric ($\Gamma_i/\Gamma_e= 1.05$) and the symmetry-broken ($\Gamma_i/\Gamma_e= 2/3$) regime, as well as at the transition point, $\Gamma_e\simeq \Gamma_i$. For all plots $\gamma/g=10^{-3}$ and $N_{\rm th}=0$ have been assumed.}
	\label{Fig2}
\end{figure}

\section{Anomalous energy transport} 
We first consider the macroscopic regime $n_0\gg 1$ and $N_{\rm th}\ll n_0$, where both thermal and quantum noise effects in Eqs.~\eqref{eq:itoeq1}-\eqref{eq:itoeq3} can be neglected.  The steady state  is then described by a set of amplitudes $\alpha_\ell^0$ and in Fig.~\ref{Fig2}(a) we plot the corresponding current $\langle J\rangle$ for $N=10$ sites. We see that transport in this system is very different from Ohm's law, but also from the ballistic flow of heat through a coupled chain of harmonic oscillators~\cite{Lin2011,Asadian2013,Bermudez2013}. Overall, we find regimes of normal transport, where for fixed injection rate $\Gamma_i$ the current increases with increasing extraction rate $\Gamma_e$, but also regimes of anomalous transport, where the opposite dependence is observed. 

For $\Gamma_i<g$ there is a range of rates $\Gamma_e$ where the current is completely stalled and only re-establishes at higher extraction rates. This counterintuitive behavior~\cite{Peng2014,Brandstetter2014} can be traced back to the fact that within this parameter range all eigenvalues of the linear chain, i.e., when saturation effects are neglected, have a negative real part and the whole network is damped to zero. 
In all other parameter regimes the analysis of the linear chain predicts amplified solutions with a maximal gain rate that simply increases with increasing $\Gamma_i$, see Appendix~\ref{app:Linear}. This behavior of the linear chain is not at all reflected in the stationary current shown in Fig.~\ref{Fig2}(a), 
which instead has a sharp maximum around $\Gamma_e \simeq \Gamma_i$. For $\Gamma_e=\Gamma_i$ the current then saturates at  $\langle J\rangle\simeq g n_0$ above $\Gamma_i=\Gamma_e=4g$ as shown in Fig.~\ref{Fig2}(b). Note that for  $\gamma\rightarrow 0$ the current can exhibit sharp discontinuities near this symmetry line, where it  jumps abruptly within a range $\delta \Gamma_e\sim \mathcal{O}(\gamma)$.  At high rates, $\Gamma_i/g>4$,  also a bistable regime exists, where the current depends on the order in which the rates are switched on. However, in our analysis below we find that these fine-tuned features are washed out in the presence of noise and therefore they are less relevant for understanding transport in the microscopic regime.

\section{Symmetry-breaking}

In Fig.~\ref{Fig2}(c) we also plot the occupation numbers $|\alpha_\ell^0|^2$, i.e., the stationary distribution of the energy along the channel.  In contrast to conventional transport scenarios, where the energy distribution is flat or monotonically decreasing~\cite{Lin2011, Asadian2013,Bermudez2013}, here the chain exhibits an alternating zig-zag structure. For  $\gamma\rightarrow 0$ and $N$ even we obtain (see Appendix~\ref{app:Amplitudes})
\begin{equation}\label{eq:Amplitudes}
|\alpha^0_{\ell}|^2=\left| A \sin(k_0 \ell)+ B \cos (k_0 \ell)\right|^2,
\end{equation}
where $k_0=\pi(N+2)/(2N+2)$ for  $\Gamma_{i}>\Gamma_e$ and  $k_0=\pi/2$ for $\Gamma_{i}<\Gamma_e$. Eq.~\eqref{eq:Amplitudes} shows that the stationary current is carried by a single mode with wavevector $k_0\approx \pi/2$, which is the mode supporting the highest current. However, since the saturable absorber can only extract a finite amount of energy per unit of time, most of the energy current is reflected at the extraction site and forms a standing wave.

While a standing-wave pattern is observed in all parameter regimes, the boundary conditions depend on the relation between $\Gamma_i$ and $\Gamma_e$. For $\Gamma_{i}>\Gamma_e$ the two ends of the chain have exactly the same amplitude, $|\alpha_1^0|^2\simeq |\alpha_N^0|^2\simeq |A|^2$ and $|B/A|\ll 1$. In contrast, for $\Gamma_{i}<\Gamma_e$ the amplitude of the gain mode is much higher than the amplitude of the loss mode, $|\alpha_1^0|^2\gg|\alpha_N^0|^2$. Therefore, for $\Gamma_i>\Gamma_e$ the steady-state energy distribution of this network exhibits a left-right (parity) symmetry, which is broken above the transition point $\Gamma_e\simeq\Gamma_i$. Exactly at this point we obtain $B\simeq -i A$ and the transport becomes unidirectional, $\alpha^0_{\ell}\sim e^{ik_0\ell}$. Note that also this behavior of the steady-state amplitudes cannot be derived by looking at the mode function of the most unstable mode of the linear chain. This mode always has the highest amplitude on site $\ell=1$, such that gain is maximized. A more detailed derivation and discussion of the steady-state amplitudes is given in Appendix~\ref{app:Amplitudes}.

The breaking of a spatial symmetry in systems with gain and loss is reminiscent of the effect of $\mathcal{PT}$-symmetry breaking \cite{Ganainy2018} in systems with equal gain and loss rates. 
Interestingly, in the current system such a symmetry is not present in the underlying equations of motions, since for  $\Gamma_i\neq \Gamma_e$ the oscillators at the injection and extraction sites evolve with very different rates. Additional numerical and analytical results summarized  in Appendix~\ref{app:Universality} show that the emergence of a symmetric stationary phase also does not rely on the specific choice of the cutoff function $f(x)$ and can be found even in situations where the gain and loss processes saturate at different amplitudes, $n_{0}^{(1)}\neq n_{0}^{(N)}$. For such general cases, the symmetric phase is established as long as there is an amplitude $\alpha^0=\alpha_1^0=\alpha_N^0$ such that energy conservation 
 \begin{equation}
 \Gamma_{i}(\alpha^0)-\Gamma_e(\alpha^0)\simeq \frac{\gamma N}{2}
 \end{equation} 
 can be satisfied. Only for larger chains this symmetry degrades when either $N\gamma > g$ or a finite amount of disorder prevents a free propagation of excitations (see Appendix~\ref{app:Disorder}).
Therefore, consistent with previous observations in specific two-mode systems~\cite{Hassan2015,Ge2016}, we find that the emergence of steady-state symmetries and the breaking thereof is a generic  mechanism in active oscillator networks. In this context, the so-called $\mathcal{PT}$-symmetric configuration, $\Gamma_i= \Gamma_e$, appears naturally as the phase boundary, along which additional symmetry-breaking transitions can take place~\cite{Kepesidis2016}.

\begin{figure}
	\centering
	\includegraphics[width=\columnwidth]{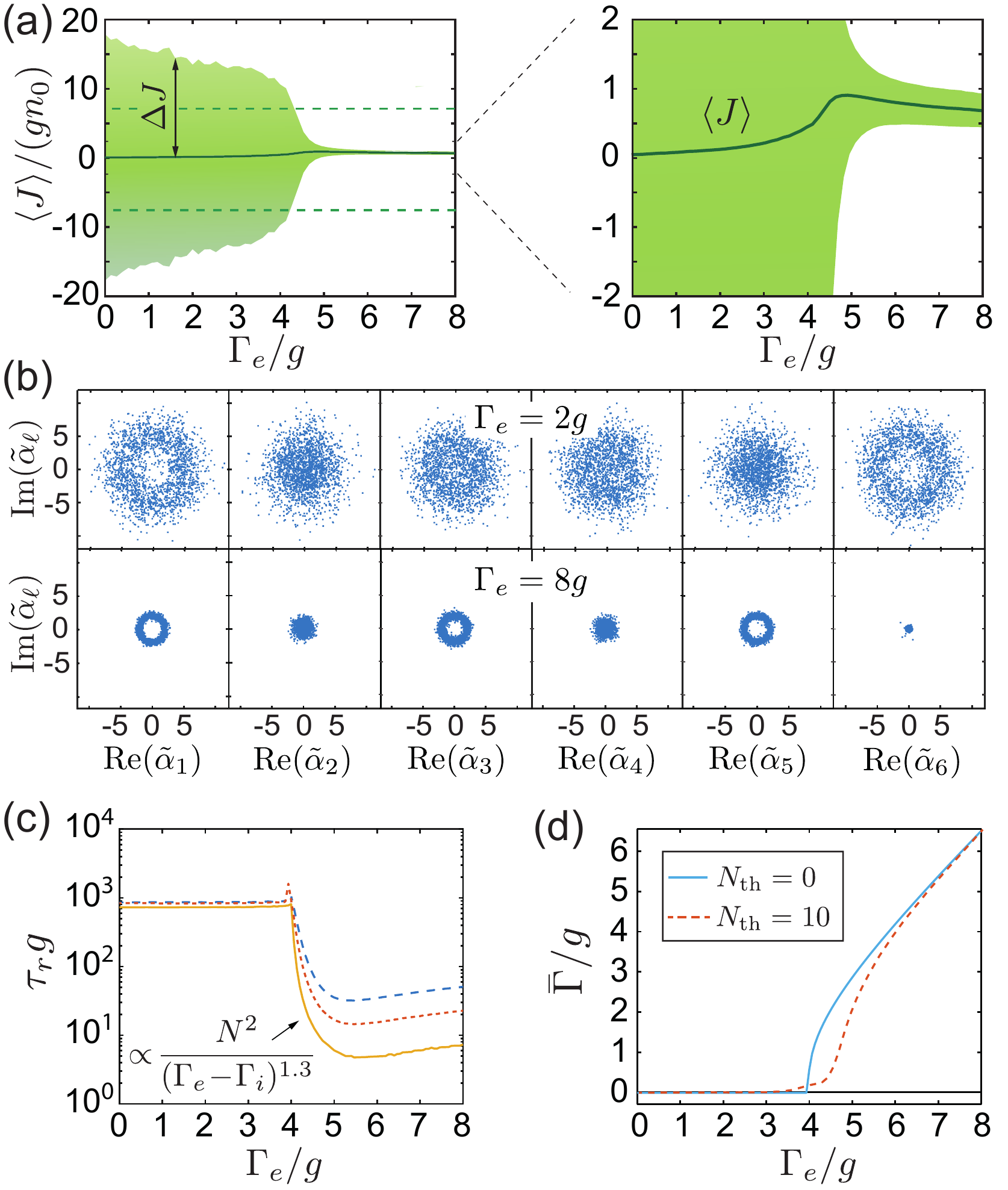}
	\caption{(a) The average current $\erw{J}$ (solid line) and the current fluctuations $\Delta J$ (shaded area) are plotted for a chain of  $N=6$ oscillators coupled to a thermal environment with $N_{\rm th}/n_{0}=10$. The dashed lines indicate the range of current fluctuations in thermal equilibrium. The marginal phase space distributions $P_\ell(\alpha_\ell)$ are shown in (b) in the symmetric ($\Gamma_{e}=2g$) and in the symmetry-broken ($\Gamma_{e}=8g$) regime.  Here $\tilde \alpha_\ell=\alpha_\ell/\sqrt{n_0}$. (c)  Plot of the relaxation time $\tau_r$ as a function of $\Gamma_{e}$ for $N=4,6,8$ oscillators and $N_{\rm th}=0$. (d) The average dissipation rate $\bar \Gamma=\erw{\erw{\Gamma_{e}(\alpha_{N})-\Gamma_{i}(\alpha_{1})}}$ in the absence and presence of thermal noise.
	In all plots fixed values of $\Gamma_{i}/g=4$ and $\gamma/g=10^{-3}$ have been assumed.}
	\label{Fig3}
\end{figure}

\section{Current fluctuations}\label{sec:Thermal}
To understand the  consequences of this symmetry-breaking transition for microscopic transport, we consider now the thermal regime, $n_0\gg1$ and  $N_{\rm th} \sim n_0$. In this case, quantum effects are still small, but noise from the environment can no longer be neglected and induces strong fluctuations of the current,  $\Delta J=\sqrt{\langle J^2\rangle-\langle J\rangle^2}$. In equilibrium, $\Delta J=g N_{\rm th}/\sqrt{2}$ and for $\Delta J/\langle J\rangle\sim N_{\rm th}/n_0>1$ these thermal  fluctuations exceed the average currents discussed above. In Fig.~\ref{Fig3}(a) we consider this high-noise regime and plot $\langle J\rangle $ and $\Delta J$ for $N_{\rm th}/n_0=10$, $\Gamma_i/g=4$ and for varying $\Gamma_{e}$. We see that in the symmetric phase  transport is indeed dominated by fluctuations, which even exceed the thermal level. This behavior changes abruptly after the transition point $\Gamma_e\simeq \Gamma_i$, beyond which a well-defined current \emph{below} the thermal noise level is established. 
This transition is also clearly visible in the steady-state distributions of the individual oscillators, $P_\ell(\alpha_\ell)$, shown in Fig.~\ref{Fig3}(b). For $\Gamma_{i}>\Gamma_{e}$ we observe strong fluctuations, but the distributions are still symmetric with respect to the center of the chain, i.e., $P_\ell\simeq P_{N-\ell+1}$.  For $\Gamma_{e}>\Gamma_{i}$ this symmetry is broken and fluctuations are strongly suppressed.

The striking difference in the current noise can be related to an equivalent change in the response of the network. In Fig.~\ref{Fig3}(c) we plot the relaxation time $\tau_r$,  i.e., the time it takes for the amplitude $\alpha_1$ to relax back to its steady-state value after a small perturbation has been applied to site $\ell=1$. For details about the numerical procedure that has been used to determine $\tau_r$, see Appendix~\ref{app:Numerics}.  In the symmetric phase this time constant is approximately  independent of $\Gamma_i$, $\Gamma_e$ and $N$. It is essentially determined by the bare damping rate, $\tau_r\sim \gamma^{-1}$, and diverges in the limit $\gamma\rightarrow0$. In the symmetry-broken phase a much faster response, $\tau_r\sim O(\Gamma_e^{-1})\sim N^2$ is observed. At the transition point the relaxation time diverges as $\tau_r\sim (\Gamma_e-\Gamma_i)^{-\xi}$, where we find $\xi\simeq 1.3$ from numerical simulations. This behavior is very different from a laser or from other non-equilibrium phase transitions, where the relaxation time diverges only at the transition point, but is finite and of similar magnitude in both phases~\cite{Diehl2010,Nagy2011,Oztop2012,Kessler2012,Casteels2016,Hwang2018}. As shown in Fig.~\ref{Fig3}(c), in the current system the relaxation time diverges (in the limit $\gamma\rightarrow 0$) within the whole symmetric phase. 

To provide a connection between the symmetry of $P(\{\alpha_\ell\})$ and the current noise, it is useful to consider the mean damping rate $\bar \Gamma=\erw{\erw{\Gamma_{e}(\alpha_{N})-\Gamma_{i}(\alpha_{1})}}$~\cite{Kepesidis2016}, i.e., the average difference between energy injection and extraction rates. Due to the symmetry of the marginal distributions, this rate is vanishing small in the symmetric phase, $\bar \Gamma\sim O(\gamma)$ [see Fig.~\ref{Fig3}(d)]. By breaking this symmetry, a finite value $\bar \Gamma\gg \gamma$ is established for $\Gamma_e>\Gamma_i$. This then leads---on average---to an efficient cooling of fluctuations and the possibility for subthermal energy transport. Again this behavior shows a close analogy to conventional $\mathcal{PT}$-symmetric systems~\cite{Ganainy2018}. In such systems the breaking of the parity symmetry of the eigenstates of a non-Hermitain matrix is accompanied by a transition from real to imaginary eigenvalues, i.e. a transition from a purely oscillatory to an exponentially damped or amplified dynamics~\cite{Ganainy2018}.  The order parameter $\bar \Gamma$ generalizes this effect to steady-state distributions of nonlinear gain-loss systems~\cite{Kepesidis2016}, where the conventional definition of $\mathcal{PT}$-symmetry breaking is no longer meaningful. Note that  the scaling of $\bar \Gamma$ near the transition point is not related to the branching of eigenvalues near an exceptional point~\cite{Ganainy2018} and depends solely on the saturation function $f(a^\dag a)$. This is illustrated by additional numerical examples presented in Appendix~\ref{app:Universality}.   Importantly, all these examples show that also the characteristic cancellation of the average dissipation rate is a much more general effect and occurs as well in gain-loss systems where no symmetry is present on a fundamental level.


\section{Quantum noise limit} 

 \begin{figure}
	\centering
	\includegraphics[width=\columnwidth]{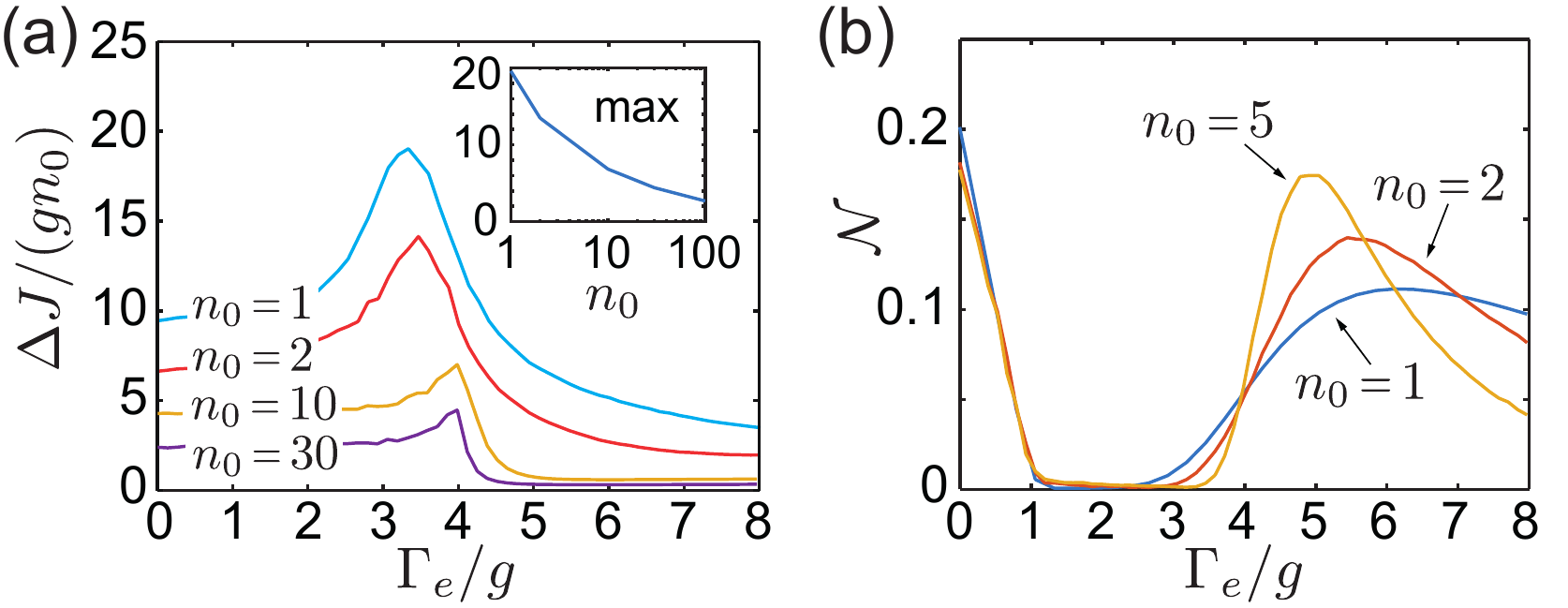}
	\caption{(a) Plot of the current fluctuations $\Delta J$ for $N=2$ oscillators in the quantum noise limit,  $N_{\rm th}=0$, and for different saturation numbers $n_{0}=1,2,10,30$.  The inset shows the scaling of the maximum of the fluctuation peak as a function of $n_0$.  (b) Entanglement negativity $\mathcal{N}$~\cite{Zyczkowski,Vidal} of the steady-state density operator for $n_{0}=1,2,5$. For these plots a fixed injection rate $\Gamma_{i}/g=4 $ and a bare damping rate of (a) $\gamma/g=10^{-3}$ and (b) $\gamma/g=10^{-2}$ have been assumed. The results in this figure have been obtained from the semiclassical stochastic differential equations~\eqref{eq:itoeq1}-\eqref{eq:itoeq3} for $n_0\geq 10$ and from stochastic wavefunction simulations of the full density operator for $n_0=1,2,5$. See Appendix~\ref{app:Numerics} for more details about the numerical simulations.}
	\label{Fig4}
\end{figure}

From Eq.~\eqref{eq:itoeq1} we see that even for $N_{\rm th}\approx 0$, the network is still affected by quantum noise $\sim\sqrt{\Gamma_i(\alpha_1)}\xi_1(t)$. In the regime $N_q=\Gamma_i/\gamma>N_{\rm th}$, this noise dominates over thermal fluctuations and represents a fundamental limitation for energy transport deep in the quantum regime, $n_0\sim O(1)$. Fig.~\ref{Fig4}(a) shows that for $n_0\gg 1$ the sharp transition between a noisy and a coherent transport regime still prevails, even for $N_{\rm th}=0$. As the saturation number $n_0$ is lowered, the relative level of fluctuations increases, develops a peak at the transition point and becomes much more pronounced also in the symmetry-broken phase. Note that for small $n_0\lesssim 10$ the mapping of the master equation onto a Fokker-Planck equation is no longer valid and a full simulation of Eq.~\eqref{eq:mastereq} must be performed (see Appendix~\ref{app:Numerics}). Therefore, due to the large Hilbert space and large separation of time scales involved in such simulations, the results in Fig.~\ref{Fig4} are restricted to $N=2$ oscillators. 

Access to the full density operator also allows us to investigate true non-classical quantities, such as the entanglement established between the injection and extraction sites. As shown in  Fig.~\ref{Fig4}(b) for different $n_{0}=1,2,5$, a significant amount of entanglement exists for $\Gamma_{e}<g$, it then vanishes in the rest of the symmetric phase, and peaks again right after the transition point. Therefore, this plot reveals an additional substructure, which is not reflected in the mean current or its fluctuations.  This entanglement between source and drain can  be relevant for thermodynamical considerations, where not only the flow of energy, but also changes in entropy through mutual (quantum) correlations must be taken into account.  Note, however, that for a more detailed study of entanglement it is necessary to go beyond our simply model and explicitly include specific implementations of quantum generators and engines in the dynamics.

\begin{figure}
	\includegraphics[width=\columnwidth]{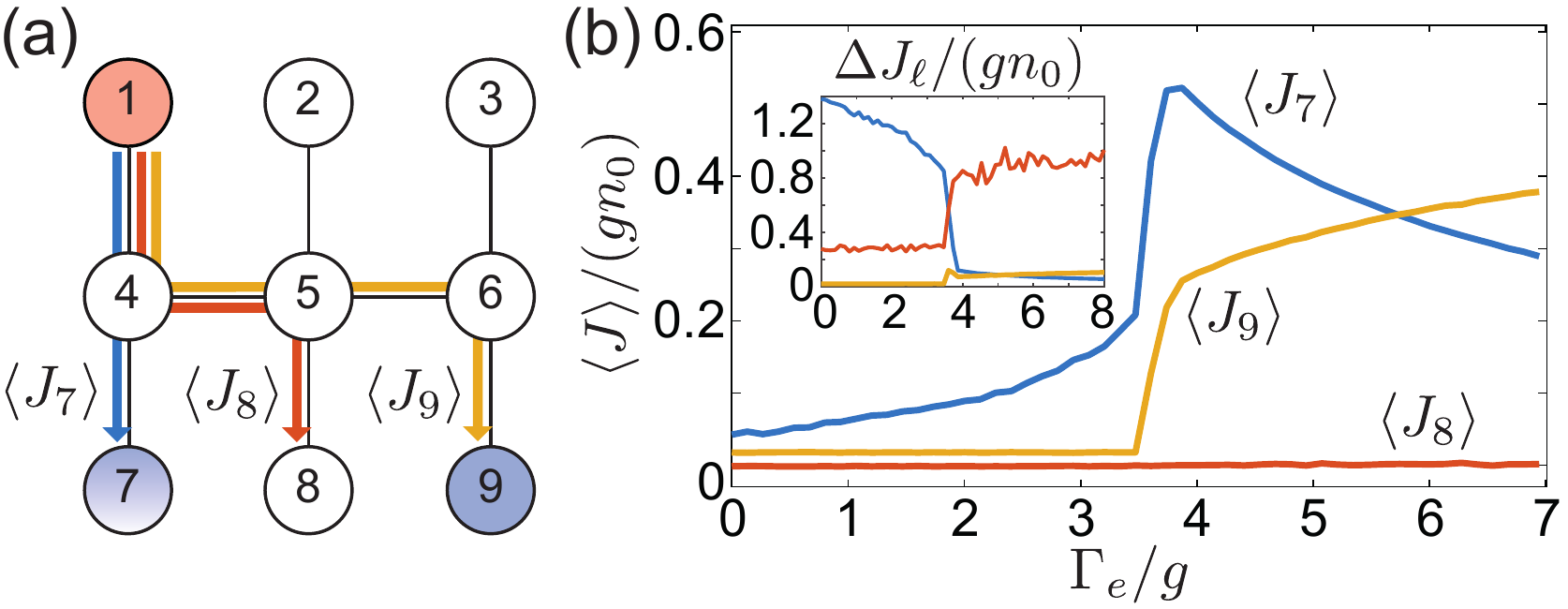}
	\caption{(a) Sketch of a 2D power grid with multiple active sites. (b) Plot of the average currents $\langle J_\ell\rangle$ flowing from site 1 to sites $\ell=7,8,9$ for fixed $\Gamma_{i}^{(1)}/g=\Gamma_{e}^{(9)}/g=4$ and varying rate $\Gamma_{e}^{(7)}$. For this plot it is assumed that all oscillators are coupled to a thermal bath with a moderate occupation number $N_{\rm th}/n_{0}=3$ and $\gamma/g=10^{-3}$. The inset shows the resulting current fluctuations. }
	\label{Fig5}
\end{figure}

\section{Discussion and conclusion}

In summary, we have shown that the transfer of  energy between two active quantum machines can be very different from what one would intuitively expect based on Ohm's or Fourier's law.  Most importantly, we have identified the emergence and breaking of spatial symmetries as a generic feature of such networks, which drastically affects their dynamical response and noise properties. This aspect is of particular relevance in the microscopic regime where thermal and quantum fluctuations are unavoidable and play a dominant role.

The transport effects analyzed here in detail for a single channel will be highly relevant as well for understanding the flow of energy through more complex networks with multiple active sites. To illustrate this point, we consider in Fig.~\ref{Fig5}(a) a small multi-port network where energy is extracted at two sites 7 and 9 with rates  $\Gamma_{e}^{(7)}$ and $\Gamma_{e}^{(9)}$, respectively. Fig.~\ref{Fig5}(b) shows that although $\Gamma_{e}^{(9)}\gg\Gamma_e^{(7)}$, only a residual thermal current is initially flowing from site 1 to site 9. This counterintuitive behavior can be explained by the fact that a symmetric standing wave is formed between sites 1 and 7, which results in a vanishing amplitude $\alpha_4\approx 0$ at the crossing site. Once $\Gamma_{e}^{(7)}$ is increased above the value of about $\Gamma_i^{(1)}$, the symmetry breaks and $\alpha_4\neq0$ now supports a large current flowing to site 9. As a result, we obtain a transistor-like behavior, where a small increase of losses in one site leads to a sudden increase of the energy current through another part of the network. Simultaneously, there are sharp jumps in the level of current fluctuations, in analogy to what we have found above for the 1D chain.

This brief outlook already shows that the combination of interference, nonlinear symmetry-breaking effects and the prominent role of noise makes the operation of microscopic power grids a very rich and complex problem, which is still little understood. The current analysis reveals the important part in this problem that is played by quantum fluctuations as a fundamental source of noise as well as by the topology of the network, which determines whether these fluctuations are enhanced or suppressed.  

\section{Acknowledgement} We thank Alexander Carmele, Wolfgang Niedenzu, Stefan Rotter and Henning Schomerus for stimulating discussions. This work was supported by the Austrian Science Fund (FWF) through the SFB FoQuS, Grant No. F40, the START Grant No. Y 591-N16, and the DK CoQuS, Grant No. W 1210.   J.H. is a recipient of a DOC Fellowship of the Austrian Academy of Sciences (\"OAW).

\appendix

\section{Fokker-Planck equation}\label{app:FP}
In the semiclassical regime $n_{0}\gg 1$ the cutoff function $f(x)$ in Eq.~(\ref{eq:f}) varies slowly on the scale of individual excitations and the master equation can be mapped onto a Fokker-Planck equation for the Glauber-Sudarshan P-representation~\cite{Sudarhan,WallsMilburn,GardinerZoller}. This distribution function is defined by
\begin{equation}
\rho= \int \prod_\ell d^2\alpha_\ell  \, P(\{\alpha_\ell\}) \ket{\{\alpha_\ell\}}\bra{\{\alpha_\ell\}},
\label{eq:Prepresentation}
\end{equation}
where $|\{\alpha_\ell\}\rangle$ denotes a multi-component coherent state. By using the usual substitution rules~\cite{WallsMilburn,GardinerZoller}
\begin{equation}
\begin{split}
a_\ell \rho\,\, \rightarrow \,\, \alpha_\ell P, \qquad a^\dag_\ell \rho\,\, \rightarrow \,\, \left (\alpha_\ell^{*}-\frac{\partial}{\partial \alpha_\ell} \right)  P,\\
\rho a^\dag_\ell \,\, \rightarrow \,\, \alpha^*_\ell P, \qquad  \rho a_\ell \,\, \rightarrow \,\, \left (\alpha_\ell-\frac{\partial}{\partial \alpha^*_\ell} \right)  P,
 \end{split}
\end{equation} 
we can convert Eq.~\eqref{eq:mastereq} for the density operator $\rho$ into a partial differential equation for $P(\{\alpha_\ell\})$. We obtain
\begin{equation}
\frac{\partial P}{\partial t} =  \left.\frac{\partial P}{\partial t}\right|_{\rm lin} + \left.\frac{\partial P}{\partial t}\right|_{\rm nl},
\label{eq:fokkerplanklinearnonlinear}
\end{equation}
where the first term,
\begin{equation}
\begin{split}
\left.\frac{\partial P}{\partial t}\right|_{\rm lin}  =\frac{1}{2}\left[-i g \left ( \sum_{\ell=1}^{N-1} \frac{\partial}{\partial \alpha_{\ell}} \alpha_{\ell+1}  \right.  +\sum_{\ell=2}^{N}  \frac{\partial}{\partial \alpha_{\ell}} \alpha_{\ell-1}  \right )  \\ +\left. \gamma \sum_{\ell=1}^{N}  \left(\frac{\partial}{\partial \alpha_{\ell}} \alpha_\ell + N_{\rm th} \frac{\partial^{2}}{\partial \alpha_{\ell} \partial \alpha_{\ell}^{*}} \right) + c. c. \right] P,
\end{split}                                                   
\end{equation}
describes the linear chain and already has the form of a Fokker-Planck equation. 

The nonlinear dissipative terms in Eq.~\eqref{eq:mastereq} at the ends of the chain translate into higher order derivatives for the P-distribution and additional approximations are required. To do so we first use the substitution rules from above to translate the action of $f(a^{\dagger}a)$ on the density operator into a differential operator acting on a coherent state,
\begin{equation}
f(a^{\dagger}a) \ket{\alpha}\bra{\alpha} \,\, \rightarrow \,\,  \sum_{m=0}^{\infty} \bar f_{m} \alpha^m \left(\alpha^{*} + \frac{\partial}{\partial \alpha} \right)^m \ket{\alpha}\bra{\alpha}.
\end{equation}
Here the coefficients $\bar f_m$ follow from an expansion of the  operator $f(a^{\dagger}a)$ into a normally ordered series
\begin{equation}
f(a^{\dagger}a)=\sum_{m=0}^{\infty} \bar f_{m} (a^\dagger)^m a^m.
\end{equation}
By using the binomial theorem
\begin{equation}
\begin{split}
\left (\alpha^{*} + \frac{\partial}{\partial \alpha}\right)^{m}= \sum_{k=0}^{m} \binom{m}{k} (\alpha^{*})^{m-k} \frac{\partial^{k}}{\partial \alpha^{k}}\\ = \sum_{k=0}^{m} \frac{1}{k!} \frac{\partial^{k}}{\partial \alpha^{*k}} (\alpha^{*m}) \frac{\partial^{k}}{\partial \alpha^{k}},
\end{split}
\end{equation}
and integrating by parts we obtain the following substitution (omitting the site index) 
\begin{equation}
\begin{split}
f(a^{\dagger}a) \rho \,\, \rightarrow \,\, \sum_{m=0}^{\infty} \sum_{k=0}^{m} \bar f_{m} \alpha^m \frac{(-1)^k }{k!} \frac{\partial^{k}}{\partial \alpha^{*k}} (\alpha^{*m}) \frac{\partial^{k}}{\partial \alpha^{k}} P(\alpha)\\ =\sum_{k=0}^{\infty} \frac{(-1)^k }{k!} \frac{\partial^{k}}{\partial \alpha^{*k}} \left[\bar f(\alpha,\alpha^*)\right] \frac{\partial^{k}}{\partial \alpha^{k}} P(\alpha),
\end{split}
\label{eq:nlapprox}
\end{equation}
where $\bar f(\alpha,\alpha^*)=\braOket{\alpha}{f(a^\dagger a)}{\alpha}$. Since $f(a^\dagger a)$ is a function of $a/\sqrt{n_{0}}$ and $a^\dagger/\sqrt{n_{0}}$, the derivatives of $\bar f(\alpha,\alpha^*)$ scale as $\frac{\partial^{k}}{\partial \alpha^{*k}} \bar f(\alpha,\alpha^*) \propto n_{0}^{-k/2}$. Therefore, in the limit $n_{0}\rightarrow \infty$, we can neglect all derivatives and approximate 
\begin{equation}
f(a^{\dagger}a) \rho \,\, \rightarrow \,\,  \bar f(\alpha,\alpha^*) P(\alpha,\alpha^{*})+ \mathcal{O} \left ({\frac{1}{\sqrt{n_{0}}}} \right).
\label{eq:nlapprox2}
\end{equation}
Note that the definition of $\bar f(\alpha,\alpha^*)$ is based on the normally ordered series expansion and in general  $\bar f(\alpha,\alpha^*)\neq f(|\alpha|^2)$. Therefore, in our derivation we make a second approximation and neglect this difference, i.e.,
\begin{equation}
\bar f(\alpha,\alpha^*)=\braOket{\alpha}{\frac{1}{(1+a^{\dagger} a/n_{0})}}{\alpha} \approx \frac{1}{(1+\abs{\alpha}^{2}/n_{0})}.
\end{equation}
To show the validity of this approximation, we compare the function $\braOket{\alpha}{\frac{1}{(1+a^{\dagger} a/n_{0})}}{\alpha}=e^{-\abs{\alpha}^2} (-\abs{\alpha})^{-n_{0}} \left[\Gamma (n_{0},0)-\Gamma (n_{0},-\abs{\alpha}^2)\right]$ \cite{normalordering} with  the approximate form $1/(1+\abs{\alpha}^{2}/n_{0})$. Here $\Gamma(n,x)$ denotes the incomplete Gamma function. Even deep in the quantum regime, $n_{0}\approx1$,  these two expressions agree up to a few percent and become essentially identical for $n_{0}\gtrsim 10$. Therefore, we conclude that the main approximation in the derivation of our semiclassical Fokker-Planck equation arises from neglecting higher order derivatives in Eq. (\ref{eq:nlapprox}).  

Based on these considerations we obtain the following approximate substitution rules
\begin{equation}
\begin{split}
A \rho \,\, \rightarrow \,\, \frac{\alpha }{(1+\abs{\alpha}^2/n_{0})}  P(\alpha,\alpha^{*}), \\ A^\dagger \rho\,\, \rightarrow \,\, \left (\alpha^{*}-\frac{\partial}{\partial \alpha} \right) \frac{1}{(1+\abs{\alpha}^{2}/n_{0})} P(\alpha,\alpha^{*}),
\end{split}
\label{eq:Prepn1}
\end{equation}
and analogous relations for $\rho A $ and $\rho A^\dag $. All together we then obtain 
\begin{equation}
\begin{split}
\left.\frac{\partial P}{\partial t}\right|_{\rm nl} 
=\frac{1}{2}\left[ -\frac{\partial}{\partial \alpha_{1}} \Gamma_{i}(\alpha_{1}) \alpha_{1}
+ \frac{\partial^{2}}{\partial \alpha_{1} \partial \alpha_{1}^{*}}  \Gamma_i(\alpha_1) 
\right. \\
\left. + \frac{\partial}{\partial \alpha_{N}} \Gamma_{e}(\alpha_{N}) \alpha_{N}+ c.c.
\right] P,
\end{split}                                                   
\end{equation}
where $\Gamma_{i,e}(\alpha)=\Gamma_{i,e} f^2(|\alpha|^2)$. After this approximation, the resulting Fokker-Planck equation~\eqref{eq:fokkerplanklinearnonlinear} can be mapped onto the set of stochastic differential equations (3)-(5)~\cite{Gardiner}.

\section{Linear chain}\label{app:Linear}

\begin{figure}
	\includegraphics[width=\columnwidth]{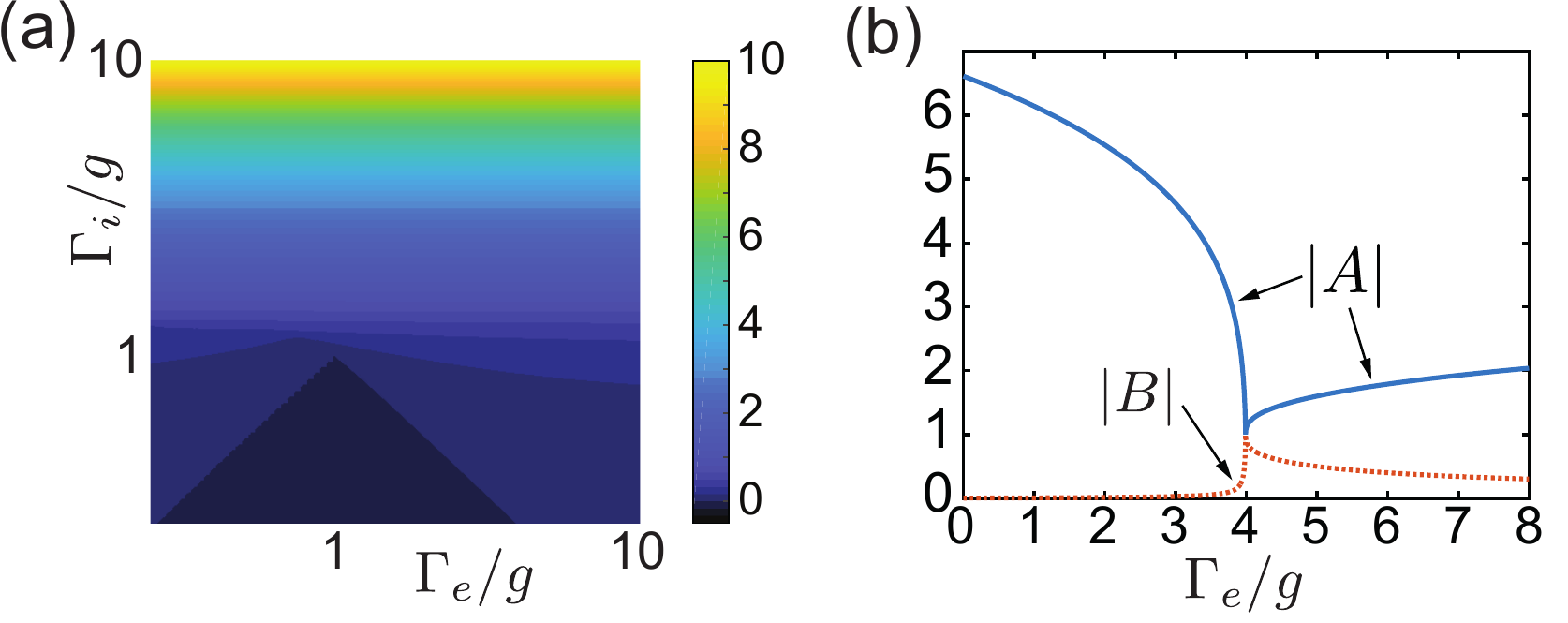}
	\caption{(a) Plot of the largest real part of all the eigenvalues of a linear chain of $N=10$ oscillators. (b) The coefficients $\abs{A}$ (solid line) and $\abs{B}$ (dotted line) used in the ansatz in Eq.~\eqref{eq:ansatz} are plotted for a fixed injection rate $\Gamma_{i}= 4 g$ and as a function of  $\Gamma_{e}$. In both plots a damping rate of $\gamma=10^{-3} g$ has been assumed. }
	\label{fig:lineareig}
\end{figure}

In Fig.~\ref{fig:lineareig}(a) we plot the largest real part of all the eigenvalues obtained from the dynamical matrix of a  linear chain where $\Gamma_{i,e}(\alpha)=\Gamma_{i,e}$. As long as all eigenvalues have a negative real part, the chain is damped to zero. This only occurs in the `stalled' phase where $\Gamma_i<g$ and $\Gamma_i \le \Gamma_e<g^2/\Gamma_i$. Otherwise, we see that the structure of the current plotted in Fig. 2(a) is not at all reflected in the eigenvalue structure of the linear chain.

\section{Steady state amplitudes}\label{app:Amplitudes}

In the limit $n_{0}\gg 1$ and $N_{\rm th}/n_{0} \rightarrow 0$, the stochastic terms in Eqs.~\eqref{eq:itoeq1}-\eqref{eq:itoeq3} can be neglected and we obtain a set of ordinary differential equations with steady-state amplitudes $\alpha_\ell^0$. To obtain analytic insights about the steady state of the chain in this regime, we consider in the following the slightly simplified scenario, where only sites $\ell=1$ and $\ell=N$ are affected by the bare decay $\gamma\ll g$, while all the other oscillators evolve coherently. 

We are interested in the long-time dynamics of the chain and make the following ansatz for the amplitudes
\begin{equation}
\alpha^0_{\ell}(t)=\sqrt{n_0} e^{-i \omega t}\left[A \sin(k_{0} \ell)+B \cos (k_{0} \ell)\right],
\label{eq:ansatz}
\end{equation}
where $A,B \in \mathbb{C}$ and $\omega=  g \cos(k_0)$. With this ansatz the current between two sites is
\begin{equation}
\langle J_\ell \rangle  = g {\rm Im}\{  (\alpha^0_{\ell-1})^* \alpha^0_{\ell} \}= g n_{0} {\rm Im}\{A B^* \} \sin(k_{0}).
\end{equation}
To obtain a steady state configuration that maximizes the energy transfer, we look for solutions with $k_{0}$ equal or close to $\pi/2$. By writing $k_0=\pi/2+\delta$ we obtain the equation
\begin{eqnarray}
\begin{split}
\left [\frac{\Gamma_{i}/2}{(1+\abs{A \cos(\delta)-B \sin(\delta)}^2)^2}-\frac{\gamma}{2} \right]\\ \times (A \cos(\delta)-B \sin(\delta)) - i \frac{g}{2} B=0,
\end{split}
\end{eqnarray}
from the equation of motion for $\alpha_1$. Similarly, from the equation of motion for $\alpha_N$ we obtain
\begin{equation}
\begin{split}
\left[-\frac{\Gamma_{e}/2}{(1+\abs{A \sin(\delta N)+B \cos(\delta N)}^2)^2}-\frac{\gamma}{2} \right] \\
\times (A \sin(\delta N)+B \cos(\delta N))\\
-i \frac{g}{2} ( A \cos(\delta (N+1))-B \sin(\delta(N+1))=0,
\end{split}
\end{equation}
for the case where $N$ is even. To proceed with our analysis we must distinguish between the symmetric ($\Gamma_i>\Gamma_e$) and the symmetry-broken regime ($\Gamma_e>\Gamma_i$) and between an even and an odd number of oscillators.  In the following we will only carry out the analysis for an even number of oscillators, however the analysis for $N$ odd can be done in an equivalent manner.


\subsection{Symmetry-broken phase}	
We first consider the regime $\Gamma_e>\Gamma_i$ and $N$ even. In this case the choice $k_0=\pi/2$ results in $\omega=0$ and a symmetry-broken solution for the amplitudes, $|\alpha_1^0|\gg |\alpha_N^0|$. 	The remaining parameters $A$ and $B$ are determined by the two coupled equations

\begin{eqnarray}
\left (\frac{\Gamma_{i}}{(1+\abs{A}^2)^2}-\gamma\right) A - i g B=0,\\
\left (\frac{-\Gamma_{e}}{(1+\abs{B}^2)^2}-\gamma \right) B - i g A=0.
\end{eqnarray}
These equations have a solution for $\Gamma_{e} \ge \Gamma_{i}$, but not for $\Gamma_{i}>\Gamma_{e}$. Although these equations can still be solved analytically, the results are already quite involved. However, sufficiently deep in the symmetry-broken phase we can neglect the bare decay $\gamma$ and approximate $\Gamma_e(B)\approx \Gamma_e$. We then obtain
\begin{equation}
\abs{A}^2\simeq \sqrt{\frac{\Gamma_{i}\Gamma_{e}}{g^2}}-1, \qquad B\simeq -i  \frac{g}{\Gamma_e } A,
\end{equation}
and the current
\begin{equation}
\langle J\rangle \simeq \frac{g^2 n_{0}}{\Gamma_e } \left (\sqrt{\frac{\Gamma_{i}\Gamma_{e}}{g^2}}-1 \right).
\end{equation}

\subsection{Symmetric phase}
For $N$ even and  $\Gamma_i>\Gamma_e$, the choice $k_0=\pi/2$ would results in an asymmetric steady state and also the resulting equations for $A$ and $B$ do not have a solution for $\Gamma_i>\Gamma_e$. To recover a symmetric solution with a maximal current we choose $\delta=\pi/(2(N+1))$. In this case the chain undergoes persistent oscillations with frequency $\omega=g \sin(\delta)$. By defining $\tilde{A}=A \cos(\delta)$ and using the approximation $B\sin(\delta) \approx 0$,  the resulting equations simplify to
\begin{eqnarray}
\left (\frac{\Gamma_{i}}{(1+\abs{\tilde{A}}^2)^2}-\gamma \right) \tilde{A} - i g B=0,\\
\left (- \frac{\Gamma_{e}}{(1+\abs{\tilde{A}}^2)^2}-\gamma \right) \tilde{A}+i g B=0. 
\end{eqnarray}
Therefore, we obtain the amplitudes
\begin{equation}
\abs{\tilde{A}}^2=\sqrt{\frac{\Gamma_{i}-\Gamma_{e}}{2 \gamma}}-1, \qquad B=-i  \frac{\gamma}{g} \frac{\Gamma_{i}+\Gamma_{e}}{\Gamma_{i}-\Gamma_{e}} \tilde{A},
\end{equation}
and, since $\sin(k_0)=\cos(\delta)$, the current
\begin{equation}
\langle J\rangle =n_{0} \gamma \frac{\Gamma_{i}+\Gamma_{e}}{\Gamma_{i}-\Gamma_{e}} \left (\sqrt{\frac{\Gamma_{i}-\Gamma_{e}}{2 \gamma}}-1 \right).
\end{equation}

\subsection{Symmetry breaking transition}	
Near the transition point we find $|A|\simeq |B|$. More precisely,  from the solution in the symmetric regime we see that $B=-iA$, at a value of
\begin{equation}
\Gamma_e^*=\Gamma_i \frac{g-\gamma}{g+\gamma}\approx \Gamma_i- \frac{2\Gamma_i }{g} \gamma.
\end{equation}
Near this parameter the standing wave turns into a running wave $\alpha_\ell^0\sim e^{ik_0\ell}$ and the current is close to maximum and scales as $\erw{J}_{\rm max} \propto \sqrt{\Gamma_{i}}$.
Although the symmetric solution exists up to  $\Gamma_e^{**}=\Gamma_i-2\gamma$, the stability analysis reveals that for $\Gamma_i>g$ the symmetric solution becomes unstable before, at around $\Gamma\simeq \Gamma_e^*$. In the regime of interest, $\gamma/g\rightarrow 0$, these differences become negligible and the transition is simply given by $\Gamma_e= \Gamma_i$. The dependence of the coefficients $A$ and $B$ around the transition point is shown in Fig.~\ref{fig:lineareig}(b).

\subsection{Damping of all oscillators}
The results derived so far for a chain without damping of the oscillators in the middle agree in essence with the results obtained for two coupled oscillators (see also Ref.~\cite{Hassan2015}). However, 
while in the symmetry-broken phase the bare damping $\gamma$ has a negligible effect, it determines the value of the current in the symmetric phase. In this regime it is thus important to analyze the steady state also for the full system, where all oscillators are weakly damped. In this case the equation 
\begin{equation}\label{eq:alphaelldot}
\dot \alpha_\ell = -\frac{\gamma}{2}\alpha_\ell +i\frac{g}{2}(\alpha_{\ell-1}+\alpha_{\ell+1}),
\end{equation}
cannot be fulfilled by the ansatz~\eqref{eq:ansatz}. However, for $\gamma/g\ll1$ the correction are small and we can still use this ansatz with the same $k_0$ as above as a first approximation. For simplicity we focus on $N$ odd where $k_{0}=\frac{\pi}{2}$. Then, summing the equations of motion for every other site we obtain
\begin{equation}
\begin{split}
&\sum_{\ell=1}^{(N-1)/2} (-1)^{\ell+1} \dot{\alpha}_{2\ell-1} =
\\&\sqrt{n_{0}} \left (\Gamma_{i}(A) A-\gamma \sum_{\ell=1}^{(N-1)/2}A- i g B \right)=0
\end{split}
\end{equation}
and for the last site,
\begin{equation}
\dot{\alpha}_{N}= \sqrt{n_{0}} \left[(-\Gamma_{e}(A)-\gamma) A+ i g B \right]=0.
\end{equation}
From this set of equations we obtain the amplitudes
\begin{equation}
\abs{A}^2=\sqrt{\frac{2(\Gamma_{i}-\Gamma_{e})}{\gamma (N+1)}}-1, \quad B=-i \frac{\gamma}{2 g} \frac{\Gamma_{e} (N-1)+2 \Gamma_{i}}{\Gamma_{i}-\Gamma_{e}} A,
\end{equation}
and the current
\begin{equation}
\erw{J}=n_{0} \frac{\gamma}{2} \frac{\Gamma_{e} (N-1)+2 \Gamma_{i}}{\Gamma_{i}-\Gamma_{e}} \left (\sqrt{\frac{2(\Gamma_{i}-\Gamma_{e})}{\gamma (N+1)}}-1 \right).
\end{equation}
Although this result was derived for $N$ odd, it is also a good approximation for $N$ even when $N>2$.

Note that near $\Gamma_e^*$ we obtain a single traveling wave. To account first order corrections due to a finite decay $\gamma/g\ll1$, we can generalize the ansatz to $\alpha_\ell^0\sim e^{ik_0\ell}e^{-\kappa \ell}$. From Eq.~\eqref{eq:alphaelldot} we then obtain $\kappa=\gamma/(2g)$. Therefore, all our analytic estimates will remain valid as long as $N\gamma\ll1$, although numerical simulations show that most of the qualitative features survive at much larger decay rates.

\section{Universality of the symmetry-breaking transition}\label{app:Universality}	
For all the results discussed in the main text we have assumed a specific cutoff function and the same saturation occupation number $n_0$ for the gain and the loss mechanism. While the precise quantitative findings will of course depend on these assumptions, we will now demonstrate with several other examples that the essential qualitative features of the symmetry-breaking transition do not depend on these details.  

\subsection{Different gain/loss mechanisms}
\begin{figure}
	\includegraphics[width=1\columnwidth]{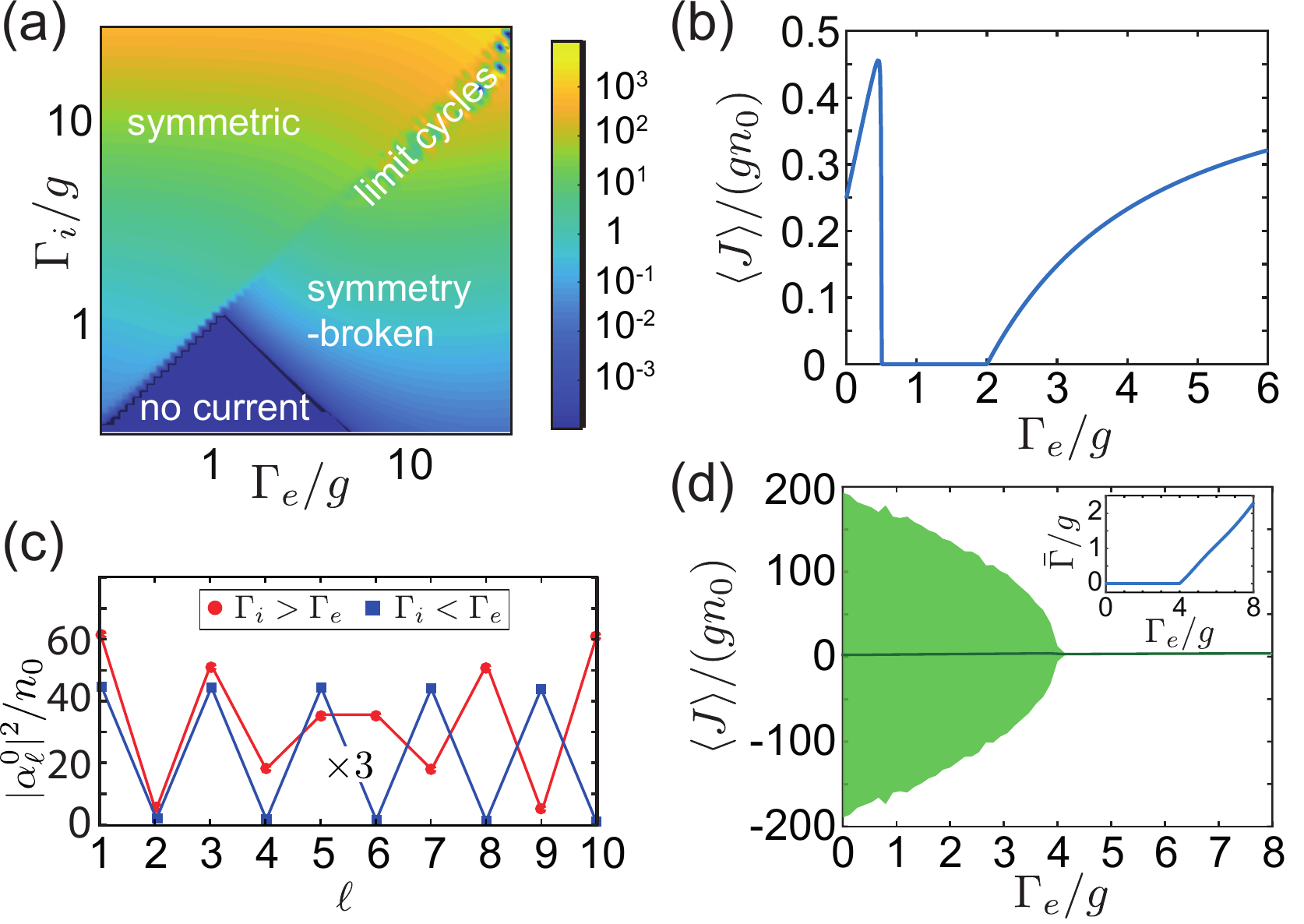}
	\caption{Energy current and symmetry-breaking for a chain of $N=10$ oscillators and  for the case where the saturation function is of the form $f(a^\dag a)=1/(1+a^\dag a/n_0)^{1/2}$. (a) Plot of the normalized steady-state energy current $\erw{J}/(gn_0)$ as a function of  $\Gamma_{i}$ and $\Gamma_{e}$ and for a damping rate $\gamma=10^{-2} g$. (b) Variation of the current as a function of $\Gamma_{e}$ for fixed injection rate $\Gamma_i/g= 0.5$ and $\gamma=10^{-3} g$. (c) Steady-state amplitudes of the chain for $\Gamma_{i}=4g$ and $\Gamma_{e}=0.5g$ (red) and $\Gamma_{e}=8g$ (blue). For better visibility the blue line is scaled by a factor of three. (d) Mean current (solid line) and range of current fluctuations (shaded area) for a network coupled to a thermal bath with $N_{\rm th}/n_{0}=10$. The inset shows the value of the average damping rate $\bar \Gamma$, as defined in Sec.~\ref{sec:Thermal}. For this plot $\Gamma_{i}=4g$ and $\gamma/g=10^{-3}$ have been assumed.}
	\label{fig:nu1}
\end{figure}

As there are many ways to engineer gain and loss, we first show that our findings do not depend on the precise form of the saturation function $f(x)$. In Fig.~\ref{fig:nu1} we consider the example of a cutoff function $f(a^\dag a)=1/(1+a^\dag a/n_0)^{\nu/2}$, where we have assumed $\nu=1$ to model a system with a weaker saturation dependence. This case corresponds, for example, to the saturation of a regular laser. Again we see the characteristic  structure of the current with a maximum at $\Gamma_e\simeq\Gamma_i$ and that this maximum is associated with a transition between a symmetric and a symmetry-broken energy distribution.  For $\Gamma_i\gtrsim  3.4g$ we obtain a region, where the current does not have a precise value and the whole chain settles into a limit cycle. Such a behavior has previously been predicted for a $\mathcal{PT}$-symmetric system, $\Gamma_i=\Gamma_e$, where $\nu=1$ has been identified as a special case, where no real symmetry-breaking occurs \cite{Kepesidis2016}.  However, in the presence of thermal noise [see Fig.~\ref{fig:nu1}(d)] these limit cycles are no longer visible and qualitatively we obtain the same transition between a noise-dominated and a coherent transport regime as in the main part of the paper. The same behavior is also found for stronger nonlinearities, $\nu=3$, and other saturation functions with different functional dependencies.  
 


\subsection{Different saturation numbers}
\begin{figure}
	\includegraphics[width=1\columnwidth]{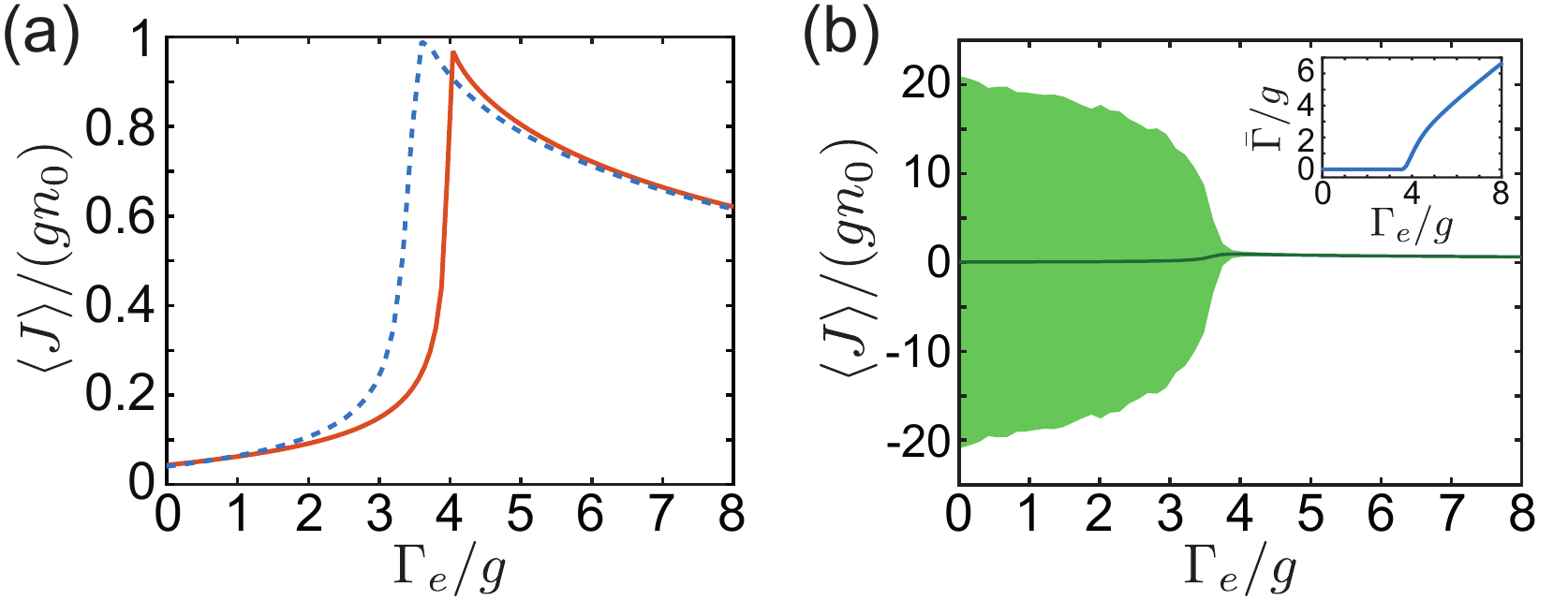}
	\caption{(a) Plot of the current for fixed injection rate  $\Gamma_{i}=4g$ and varying extraction rate $\Gamma_{e}$ for a network of $N=10$ oscillators and $N_{\rm th}=0$. The plot compares the case where the saturation numbers for gain and loss are equal, $n_{0}^{(1)}=n_{0}^{(N)}$, (red solid line) with the case where they differ by $10\%$,  $n_{0}^{(1)}=1.1 n_{0}^{(N)}$, (blue dashed line). (b) Mean current (solid line) and range of current fluctuations (shaded area) for the case $n_{0}^{(1)}=1.1 n_{0}^{(N)}$, but coupled to a thermal bath with $N_{\rm th}/n_{0}=10$. For both plots $\gamma/g=10^{-3}$ and a cutoff function as given in  Eq.~\eqref{eq:f} have been assumed.}
	\label{fig:differentn0}
\end{figure}

To further illustrate that the physical effects discussed in this work are very generic, we now return to the cutoff function given in Eq.~\eqref{eq:f}, but consider the case where the gain and the loss oscillator saturate at different amplitudes, i.e., $n_{0}^{(1)}\neq n_{0}^{(N)}$. The resulting mean currents and fluctuations are shown in Fig.~\ref{fig:differentn0}.  We see that also in this case all the qualitative features of the symmetry-breaking phase transition remain unaffected, except that the transition point is now shifted from $\Gamma_{i}=\Gamma_{e}$ 
to $\Gamma_{i} \simeq\Gamma_{e} \left(n_{0}^{(N)}/n_{0}^{(1)}\right)^2$.
%
%
Importantly, this example shows that even when different saturation mechanisms for energy injection and extraction are considered, there is still an emergent symmetric phase, which is characterized by an almost complete cancellation of the average dissipation rate $\bar \Gamma=\langle \langle \Gamma_e(\alpha_N)-\Gamma_i(\alpha_1)\rangle\rangle$.

\subsection{Conditions for a symmetric phase}
To obtain a more general result for the symmetry-breaking point, we derive a minimal condition under which a symmetric phase can exist. This condition follows from the fact that in the steady state the total energy of the system must be conserved. This means that the absorbed and dissipated energy must be the same, or
\begin{equation}
\Gamma_{i}(\alpha^0_1) |\alpha_1|^2 = \Gamma_{e}(\alpha^0_N) |\alpha^0_N|^2 +  \gamma \sum_{\ell=1}^N |\alpha^0_\ell|^2.
\end{equation}
For a symmetric state, where $|\alpha_1^0|=|\alpha_N^0|=\alpha^0$ and $|\alpha^0_\ell|^2=\eta_\ell |\alpha^0|^2$, we obtain
\begin{equation}\label{eq:EnergyConservation}
\Gamma_{i}(\alpha^0) -  \Gamma_{e}(\alpha^0) =  \gamma\mathcal{N},\qquad \mathcal{N}= \sum_{\ell=1}^N \eta_\ell.
\end{equation}
Here $\mathcal{N}=2$ for $N=2$, $\mathcal{N}\simeq (N+1)/2$ for $N$ odd and in general $\mathcal{N}\approx N/2$ for $N\gg1$. For identical cutoff functions this condition can always be satisfied by increasing the value of $\alpha^0$, as long as $\Gamma_e< \Gamma_i-\gamma \mathcal{N}$. For non-identical saturation parameters, $n_{0}^{(1)}\neq n_{0}^{(N)}$, and by approximating $f(\alpha^0)\simeq n_0^2/|\alpha^0|^4$, this argument also explains the shift of the transition point discussed above. Thus, Eq.~\eqref{eq:EnergyConservation} provides a simple minimal condition for the existence of a symmetric phase. Note, however, that for larger systems one find that for $\Gamma_i>g$ symmetry breaking already  occurs closer to the point where  $\abs{A}=\abs{B}$. For example, for $n_{0}^{(1)}= n_{0}^{(N)}$ we find the transition point approximately at
\begin{equation}
\Gamma_{e}^*=\frac{2 \Gamma_{i} (g-\gamma)}{2g+\gamma (N-1)}.
\end{equation}
As long as $\gamma N \ll g$, this result does not considerable change by changing the system size and for all results presented in the main text the transition point derived from Eq.~\eqref{eq:EnergyConservation} is a sufficient approximation.

\section{Disorder}\label{app:Disorder}

\begin{figure}
	\includegraphics[width=1\columnwidth]{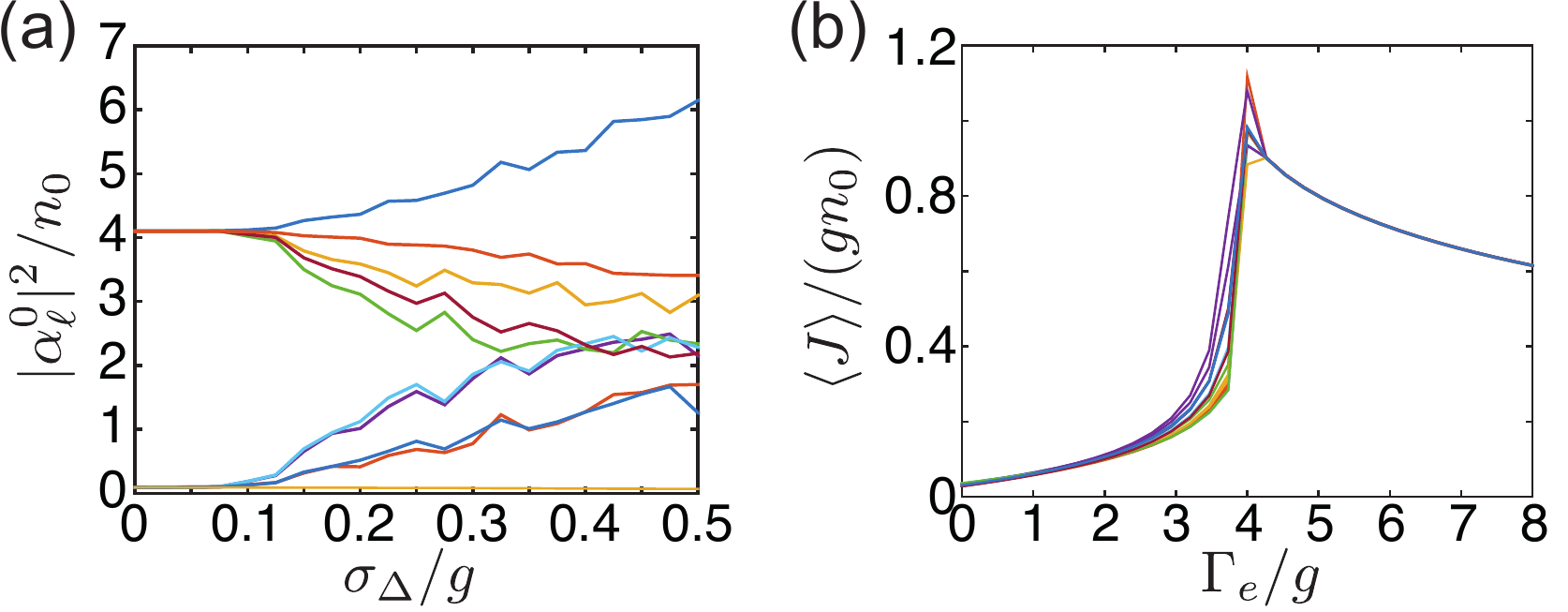}
	\caption{(a) Plot of the occupation numbers $|\alpha_\ell^0|^2$ of a chain of $N=10$ oscillators averaged over 100 realizations of random site detunings $\Delta_\ell\in[-\sigma_\Delta,\sigma_\Delta]$. (b) Current for 15 different random detuning realizations with $\sigma_{\Delta}=0.05g$. For both plots the values $\Gamma_{i}=4g$, $\Gamma_{e}=8g$ and $\gamma/g=10^{-3}$ have been assumed. }
	\label{fig:oscdisorder}
\end{figure}

For all our results presented in the main part of the paper we have considered chains of oscillators with identical frequencies $\omega_\ell=\omega_0$. To understand the robustness of the observed effects with respect to small frequency variations, which will be unavoidable in any real system, we numerically simulate the steady state of a chain of $N=10$ oscillators with frequencies  $\omega_\ell=\omega_0+\Delta_\ell$. Here the random frequency offsets for each site are chosen from a uniform distribution $\Delta_\ell\in[-\sigma_\Delta,\sigma_\Delta]$.

In Fig. \ref{fig:oscdisorder} (a) and (b) we plot the disorder-averaged steady-state occupation numbers $|\alpha^0_\ell|^2$ for each of the oscillators and the current for a few disorder realizations.  We find that for $\sigma_{\Delta}< 0.1 g$, the steady-state amplitudes reproduce almost perfectly the alternating structure predicted for the ideal case, $\Delta_{\ell}=0$.  In this regime also the current exhibits the characteristic peak structure for each individual disorder realization and is hardly affected for parameters away from the transition point. This shows that all the effects discussed in the main part of this work are insensitive  to a small amount of disorder. For $0.1 <\sigma_{\Delta}/g< 0.3$, the amplitudes still follow more or less a zig-zag structure, while for $\sigma_{\Delta}/g>0.3 $ the energy distribution is completely different from the non-detuned case and most of the energy gets localized around the gain mode. 

\section{Numerical simulations}\label{app:Numerics}
\label{sec:numerics}

For the numerical integration of the stochastic equations~\eqref{eq:itoeq1}-\eqref{eq:itoeq3} we have used the Euler Maruyama method with a time step of $\Delta t=10^{-4}/g$. For all the main plots, the stochastic equations have been vectorized and $n_{\rm traj}=50$ trajectories have been evolved simultaneously. After a time $t=5000/g \approx 5 \tau_{r}$, we have sampled $3000$ points per trajectory, separated by  $17000$ time steps, to get the steady-state distribution.  

The relaxation time in Fig. 3(c) was obtained in the absence of noise, by first determining the steady state amplitudes with high accuracy. This was implemented by a 4-th order Runge-Kutta algorithm with an accuracy of $10^{-11}$.  Then the amplitude of the gain oscillator was changed by an amount $\delta \alpha_1=1/10$ and the simulation was continued until the system has relaxed again. From the time difference $\Delta t_{r}$ between the points where the remaining occupation difference of the first oscillator was $\delta \abs{\alpha_1}^2=10^{-5}$ and $\delta \abs{\alpha_1}^2=10^{-8}$, we calculated the relaxation time as $\tau_{r}=\Delta t_{r} /\ln(10^3)$. Note that in Fig. 3(c) the relaxation rate exhibits a peak in a very small region around the transition point $\Gamma_e=\Gamma_i$, where we find almost no relaxation.  In this regime the numerically extracted values for $\tau_r$ depend on very fine details and are no longer meaningful.

For the cases $n_{0}=1,2,5$ in Fig. 4(a) and (b) we have used a stochastic quantum wavefunction method~\cite{molmer,daley} to simulate the full master equation~\eqref{eq:mastereq}. The results for $n_0=1$ were independently verified by calculating directly the steady-state density operator for a system of two coupled oscillators with $n_{\rm basis}=30$ basis states per oscillator. For the quantum trajectory simulations the evolution under the effective non-hermitian Hamiltonian has been implemented by the time evolution operator $U=e^{-i H_{\rm eff} \Delta t}$ with $\Delta t= 2\times 10^{-3}g^{-1}$, which is computed once at the beginning of the trajectory. After random times quantum jumps occur and the state gets renormalized. After $t=10000/g\approx 10 \tau_{r}$, when the system has reached the steady state, we have sampled the state after every $800$ time steps for $990000$ times to obtain the steady state density matrix. In Fig. 4(a) we used $n_{\rm basis}=70$ and $n_{\rm basis}=150$ states per oscillator for $n_{0}=1$ and $n_{0}=2$, respectively, while in Fig. 4(b) we used  $n_{\rm basis}=30,60,100$ states per oscillator for $n_{0}=1,2,3$. The entanglement negativity was obtained by calculating the $4 n_{\rm basis}$  lowest eigenvalues after partial transpose.

\end{document}